\documentstyle[12pt]{article}

\newcommand{\beq}{\begin{equation}}
\newcommand{\eeq}{\end{equation}}
\newcommand{\bea}{\begin{eqnarray}}
\newcommand{\eea}{\end{eqnarray}}
\newcommand{\tr}{\mbox{\rm Tr}}
\newcommand{\res}{\,\mbox{\rm res}}
\newcommand{\bean}{\begin{eqnarray*}}
\newcommand{\eean}{\end{eqnarray*}}
\def\cc{${\cal C}\,\,$}

\def\cl{{\cal L}}
\def\n2{{$N=2$}}
\newcommand{\reff} [1]
	{(\ref{#1})}

\renewcommand{\thefootnote}{\alph{footnote})}

\begin{document}
\thispagestyle{empty}
\bibliographystyle{plain}
\begin{flushright}ENSLAPP-L-617/96\\
solv-int/9609008\\
September 1996
\end{flushright}
\vskip 1.0truecm
\begin{center}{\bf\Large
\n2 KP and KdV hierarchies in extended superspace}
\end{center}
 \vskip 1.0truecm
\centerline{\bf F. Delduc, L. Gallot}
\vskip 1.0truecm
\begin{center} \it Laboratoire de Physique Th\'eorique ENSLAPP
\begin{footnote}{URA 14-36 du CNRS, associ\'ee  \`a l'ENS de Lyon et au LAPP\\
\hspace*{.6cm}Groupe de Lyon: ENS Lyon, 46 All\'ee d'Italie, 69364 Lyon, France
}\end{footnote}\end{center}
\vskip 2.0truecm  \nopagebreak

\begin{abstract}
We give the formulation in extended superspace of an
\n2 supersymmetric KP hierarchy using chirality preserving pseudo-differential operators. We obtain two 
quadratic hamiltonian structures, which lead to different reductions
of the KP hierarchy. In particular we find two different hierarchies 
with  the \n2
classical super-${\cal W}_n$ algebra as a hamiltonian structure.  
The relation with the formulation in $N=1$ superspace
is carried out.
\end{abstract}
\setcounter{page}{0}
\newpage
\renewcommand{\thefootnote}{\arabic{footnote})}
\setcounter{footnote}{0}
\section*{Introduction}

There has been recently an important activity in the study of \n2 supersymmetric
 hierarchies (KP \cite{popo1,aratyn,dasb1,ghosh,dasp}, generalizations of KdV 
\cite{bokris,ikrim}, Two Bosons \cite{dasb2}, NLS \cite{kriso,krisoto,dasb3}, 
etc..). The most usual tools 
in this field are the algebra of 
$N=1$ pseudo-differential operators and Gelfand-Dickey type Poisson brackets
\cite{geldik}. 
Although these systems have \n2 supersymmetry, only for very few of them 
with very low number of fields is a formulation in extended superspace known.
It is the purpose of this paper to partially fill this gap. The formalism which 
we shall present here partly originates from the article \cite{delma}.
 
It turns out that in order to construct the Lax operators of 
\n2 supersymmetric hierarchies,
one should not use the whole algebra of \n2 pseudo-differential operators, but 
rather the subalgebra of pseudo-differential operators preserving chirality. 
These operators were first considered in \cite{popo3}. 
They will be defined in section \ref{main}, where we also study the KP Lax
equations and the two associated Hamiltonian structures. It turns out that the 
first (linear) bracket 
is associated with a non-antisymmetric $r$ matrix
\cite{semenov}. Because of that, the second
(quadratic) bracket is not of pure Gelfand-Dickey type. The main result of this 
paper is that we find two
possibilities for this quadratic bracket. In fact, we show that there 
is an invertible map in the KP phase space which sends one of the 
quadratic Poisson structure into the other. However, this map does 
not preserve the Hamiltonians. 

In section \ref{reduc}, we study the possible reductions of
the KP hierarchy by looking for Poisson subspaces in the phase space. These 
are different depending on the quadratic bracket which is used.
Among these reductions, there are two different hierarchies with  the \n2
classical super-${\cal W}_n$ algebra \cite{lupope} as a hamiltonian structure.
In particular, two of the three known \n2 supersymmetric extensions 
of the KdV hierarchy \cite{Mathieu1} are found. 
They correspond to $a=-2$ and $a=4$ in 
the classification of Mathieu. These and some other examples are described in section \ref{examples}. Notice that from the known cases with a low number of fields \cite{Mathieu1,math2,popo2,yung1,Ivanov1,yung2}, 
one expects for any $n$ three hierarchies with 
super-${\cal W}_n$ as a hamiltonian structure. So our construction does not exhaust the possible cases.

We also found two hierarchies which Poisson structure 
is the classical ``small" $N=4$ superconformal algebra. In one case the 
evolution equations are $N=4$ supersymmetric, while in the other they 
are only \n2 supersymmetric. 
Finally, in section \ref{n1susy} we give the relation of our formulation with the usual formulation of the \n2 supersymmetric KP Lax equations in $N=1$ superspace \cite{inami,dasb1,dasp}. 

\setcounter{equation}{0}
\section{N=2 KP hierarchy \label{main}}

\paragraph{\n2 supersymmetry}

We shall consider an \n2 superspace with space coordinate $x$ and two
Grassmann coordinates $\theta$, $\bar\theta$. We shall use the 
notation ${\underline x}$ for the triple of coordinates 
$(x,\theta,\bar\theta)$. The supersymmetric covariant derivatives 
are defined by
\begin{equation}
\partial\equiv{\partial\over\partial x},\,\,D={\partial\over\partial\theta}
+\bar\theta\partial,\,\,\bar D={\partial\over\partial\bar\theta}
+\theta\partial, D^2=\bar D^2=0,\,\,\{ D,\bar D\}=\partial
\label{n2alg}\end{equation}
Beside ordinary superfields $H({\underline x})$ depending 
arbitrarily on Grassmann coordinates, one can also define chiral
superfields $\varphi({\underline x})$ satisfying
$D\varphi =0$ and antichiral superfields $\bar\varphi({\underline x})$ 
satisfying $\bar D\bar\varphi =0$.
We define the integration over the \n2 superspace to be
\begin{equation}
\int d^3{\underline x}\, H(x,\theta,\bar\theta)= \int dx\bar DDH(x,\theta,\bar\theta)
\vert_{\theta=\bar\theta=0}.
\end{equation}

The elements of the associative algebra of \n2 pseudo-differential operators ($\Psi$DOs) are the operators
\begin{equation}
P = \sum_{i <M} ( a_{i} +b_i[D,\bar D]+\alpha_{i} D + \beta_{i} \overline{D} )\partial^{i}
\label{pdo}\end{equation}
where $a_{i}$, $b_{i}$ and $\alpha_{i}$, $\beta_{i}$ are respectively even and odd \n2 superfields.
However, this algebra is not very manageable. 
In particular, the set of strictly pseudo-differential operators ($M=0$ in\reff{pdo}) is not 
a proper subalgebra, but only a Lie subalgebra.
Also, there are  too many fields in these operators.  We expect the phase space of the 
\n2 KdV hierarchies to consist of the supercurrents of the \n2  
${\cal W}_n$ algebras.  In extended superspace, these supercurrents are bosonic superfields, 
and there is one such superfield for a given integer dimension.
But in \reff{pdo}, each power of $\partial$ corresponds to four superfields, two even ones 
of integer dimension and two odd ones of half-integer dimension. It is thus clear that one
has to restrict suitably the form of the \n2 operators. It turns out
that a possible 
restriction is to define the set $\cal C$ of 
pseudo-differential operators $L$ preserving chirality of the form
\begin{footnote}
{Operators of this type were first considered in \cite{popo3}}
\end{footnote}
\begin{equation}
L=D{\cal L}\bar D,\,\,\,\,\,\,{\cal L}= \sum_{i <M} u_{i}\partial^{i}
\label{cpdo}\end{equation}
The coefficient functions $u_i$ are bosonic \n2 superfields. These operators  satisfy 
$DL=L\bar D=0$.
The product of two chiral operators is again a
chiral operator. The explicit product rule is easily worked out
\begin{equation}
LL'= D \left( 
{\cal L}\partial {\cal L'} +(D.{\cal L})(\bar{D}.{\cal L'}) \right) \bar{D},
\end{equation}
where we have used the notation
\begin{equation}
(D.{\cal L})=\sum_{i <M}(Du_{i})\partial^{i}.
\end{equation}
Notice that $I=D \partial^{-1}\bar{D}$ is the unit of the algebra $\cal C$.
We could have used as well the algebra $\bar{\cal C}$ of $\Psi$DOs
satisfying $\bar D\bar L=\bar L D=0$. Notice that the product of an element in \cc by 
an element in $\bar{\cal C}$ vanishes. In fact \cc and $\bar{\cal C}$
are related by transposition, $L^t=-\bar D\cl^tD\in \bar{\cal C}$. 
Although the transposition leads from ${\cal C}$ to $\bar{\cal C}$, 
there exists an anti-involution which acts inside ${\cal C}$. It is 
given by
\begin{equation}
\tau(L)=DL^t\partial^{-1}\bar 
D,\,\,\,\tau(L_{1}L_{2})=\tau(L_{2})\tau(L_{1}).
\end{equation}

Notice that it does not make sense in the algebra $\cal C$ to multiply a 
$\Psi$DO by a function. However, it is possible to multiply on the left by a chiral 
function $\phi$, $D\phi=0$
\begin{equation}
\phi L= D\phi\cl\bar D={\lambda}(\phi) L,\,\,\,{\lambda}(\phi)\equiv 
D\phi\partial^{-1}\bar D,
\end{equation}
and on the right by an antichiral function $\bar\phi$, $\bar D\bar\phi=0$
\begin{equation}
L\bar\phi = D\cl\bar\phi\bar D=L{\bar\lambda}(\bar\phi),\,\,\,
{\bar\lambda}(\bar\phi)\equiv D\partial^{-1}\bar\phi\bar D.
\end{equation}

We define the residue of the pseudo-differential operator $L$ by 
$\mbox{\rm res} L=u_{-1}$ \cite{Mathieu1}. 
The residue of a commutator is a total derivative,
$\mbox{\rm res}[L,L']=D\bar\omega+\bar D\omega$. The trace of $L$ is the integral of the 
residue
\begin{equation}
\mbox{\rm Tr}L=\int d^3{\underline x}\,\mbox{\rm res}L,\,\,\,\,
\mbox{\rm Tr}[L,L']=0.
\end{equation}
$\cal C$ can be divided into two proper subalgebras 
${\cal C} = {\cal C}_{+} \oplus {\cal C}_{-}$, where $L$ is in ${\cal C}_{+}$ if $\cal L$ is a differential operator and $L$ is in ${\cal C}_{-}$ if $\cal L$ is a 
strictly pseudo-differential operator ($M=0$ in \reff{cpdo}). We shall note
\begin{equation}
L=L_++L_-, \,\,\,L_+=D{\cal L}_+\bar D\in {\cal C}_+,\,\,\, 
L_-=D{\cal L}_-\bar D\in{\cal C}_-.
\end{equation}
Here an important difference with the usual bosonic and $N=1$ cases occurs. For 
any two $\Psi$DOs $L$ and $L'$ in $\cal C$ one has $\tr(L_{-}L'_{-})= \int d^3{\underline x}\,\res(L)\,\res(L') \neq 0$. While ${\cal C}_{+}$ is an isotropic subalgebra, ${\cal C}_{-}$
is not. One important consequence of this fact is that if one defines the 
endomorphism $R$ of $\cal C$ by $R(L)={1\over 2}(L_+-L_-)$, then $R$ is a non-antisymmetric classical $r$ matrix,
\begin{equation}
\tr (R(L)L'+LR(L'))=-\int d^3{\underline x}\res L\res L'.
\end{equation}
Notice that a non-antisymmetric $r$ matrix in the context of bosonic KP Lax
equations first appeared in \cite{kuper}.

\paragraph{KP equations}

Let us now write the evolution equations of the \n2 supersymmetric KP hierarchy.
We consider operators $L=D\cl\bar D$ in $\cal C$ of the form
\begin{equation}
\cl=\partial^{n-1}+\sum_{i=1}^{\infty}V_i\partial^{n-i-1}.
\label{KPop}\end{equation}
$L$ has a unique $n$th root in $\cal C$ of the form
\begin{equation}
L^{1\over n}=D(1+\sum_{i=1}^{\infty}W_i\partial^{-i})\bar D,
\end{equation}
and we are led to consider the commuting flows (see \cite{popo3})
\begin{equation}
{\partial\over\partial t_k}L=[(L^{k\over n})_+,L]=[R(L^{k\over n}),L].
\label{KPeq}\end{equation}
There are symmetries of these equations which may be described as 
follows. Let us first introduce a chiral, Grassmann even superfield $\varphi$ which 
satisfies
\begin{equation}
{\partial\over\partial t_k}\varphi=(L^{k\over n})_+.\varphi
\label{fieq}\end{equation}
where the right-hand side is the chiral field obtained by acting with 
the differential operator $(L^{k\over n})_+$ on the field $\varphi$.
Then the transformed operator 
\begin{equation} s(L)= \lambda(\varphi^{-1})L\lambda(\varphi) 
\label{simil}\end{equation}
satisfies an evolution equation of the same form \reff{KPeq} as that 
of $L$. 

We may also consider an antichiral, Grassmann odd superfield 
$\bar\chi$ which satisfies
\begin{equation}
{\partial\over\partial t_k}\bar\chi=-(L^{k\over n})^t_+.\bar\chi
\label{chieq}\end{equation}
Then the transformed operator 
\begin{equation} \sigma(L)=(-1)^n\lambda((D\bar\chi)^{-1})\tau(L)\lambda(D\bar\chi)
\label{chitra}\end{equation}
satisfies an evolution equation of the same form \reff{KPeq} as that 
of $L$, with the direction of time reversed.

\paragraph{Poisson brackets}

The Lax equations \reff{KPeq} are bi-hamiltonian with respect to two compatible Poisson brackets
which we now exhibit.  
Let $X$  be some $\Psi$DO in $\cal C$ with coefficients independent of
the phase space fields $\{V_i\}$, then define the linear functional 
$l_{X}(L) = \tr(LX)$. The generalization of the first Gelfand-Dickey bracket
is obvious and reads
\begin{equation}
\{ l_{X},l_{Y} \}_{(1)} (L) = \tr \left( L[X_{+},Y_{+}]-L[X_{-},Y_{-}] \right).
\label{pb1}\end{equation}
This is nothing but the linear bracket associated with the matrix $R$.

Now we turn to the construction of the second bracket. It will turn out
more complicated than the standard Gelfand-Dickey bracket because of the non-antisymmetry of the $r$ matrix. An analogous situation in the bosonic case is studied in \cite{Oevel}. We finally found two different 
possibilities.  
In order to write them down, we need to be able to separate the residue of a $\Psi$DO in 
$\cal C$ into  a chiral and an antichiral part. For an arbitrary 
superfield $H({\underline x})$, we define
\begin{equation}
H=\Phi[H]+\bar\Phi[H],\,\,\, D\Phi[H]=0,\,\,\, \bar D\bar\Phi[H]=0.
\end{equation}
This is not a local operation in $\cal C$. An explicit form may be 
chosen as
\begin{equation}
\Phi[H]=D\bar D\int d^{3}{\underline x}'\Delta({\underline 
x}-{\underline x}')H({\underline x}'),\,\,
\bar \Phi[H]=\bar DD\int d^{3}{\underline x}'\Delta({\underline 
x}-{\underline x}')H({\underline x}'),
\label{chipro}\end{equation}
where $\Delta$ is the distribution
\begin{eqnarray}
&\Delta({\underline x}-{\underline x}')= 
(\theta-\theta')(\bar\theta-\bar\theta')\epsilon(x-x'),&\label{distri}\\ 
&\partial\epsilon(x-x')=\delta(x-x'),\,\,\, \epsilon(x-x')=-\epsilon(x'-x).
\nonumber\end{eqnarray}

In the following, we shall use the short-hand notations 
$\Phi[\res[L,X]]=\Phi_X$, $\bar\Phi[\res[L,X]]=\bar\Phi_X$.
In general, $\Phi_{X}$ will not satisfy the same boundary conditions as the 
phase space fields do. However, we noted earlier that in the case of a commutator, 
the residue is a total derivative, 
$\res [L,X]=D\bar\omega+\bar D\omega$.
Here $\omega$ and $\bar\omega$ are differential polynomials in the fields. Then
one easily shows that $\Phi_{X}=D\bar\omega+\alpha$, 
$\bar\Phi_{X}=\bar D\omega-\alpha$,
where $\alpha$ is a constant reflecting the arbitrariness in the 
definition of $\Phi$, $\bar\Phi$. Up to this constant, $\Phi_{X}$ will
respect the boundary conditions. 

We are now in a position to write the two possibilities for the second bracket 
as
\begin{footnote}
{The Poisson brackets (\ref{pb2p},\ref{pb2m}) may be put in the 
general form introduced in \cite{Maillet}
\begin{equation}
\{ l_{X},l_{Y} \}_{(2)}^{a,b} (L) =\tr \left( LXa(LY)+XLb(LY)-LXc(YL)-XLd(YL)
\right)\end{equation}
However, the price to pay is that $a$, $b$, $c$, $d$ are non-local endomorphisms of $\cal C$. As an example, for the first quadratic bracket one finds
\begin{equation}
a(X)={1\over 2}(X_++\lambda(\Phi[\res X]))-{1\over 2}
(X_--\lambda(\Phi[\res X])),\,\,\, b(X)=\bar\lambda(\bar\Phi[\res X]).
\end{equation}
One easily checks in particular that $a$ is a non-local antisymmetric $r$ matrix.}
\end{footnote}
 \begin{equation}
\{ l_{X},l_{Y} \}_{(2)}^a (L) =\tr \left( LX(LY)_{+}-XL(YL)_{+}+
\Phi_Y LX+XL\bar\Phi_Y\right),
\label{pb2p}\end{equation}
and
\begin{equation}
\{ l_{X},l_{Y} \}_{(2)}^b (L) =\tr \left( LX(LY)_{+}-XL(YL)_{+}+
\Phi_Y XL+LX\bar\Phi_Y\right).
\label{pb2m}\end{equation}
These expressions do not depend on the arbitrary constant $\alpha$. 
Checking the antisymmetry 
of the Poisson brackets and the Jacobi identity can be done with 
a little effort. As usual,
the first bracket is a linearization of the two quadratic ones, that is to say
\begin{equation}
\{ l_{X},l_{Y} \}_{(2)}^{a,b} (L+zD\partial^{-1}\bar D)
=\{ l_{X},l_{Y} \}_{(2)}^{a,b} (L)
+z\{ l_{X},l_{Y} \}_{(1)} (L),
\end{equation}
and the linear bracket is compatible with each of the two quadratic brackets. 

Introducing 
the hamiltonians
${\cal H}_{k} = {n\over k}\tr(L^{k\over n})$, the KP evolution equations \reff{KPeq}
may be written as
\begin{equation}
\partial_{t_k}  \left( l_{X}(L) \right) 
= \{ l_{X},{\cal H}_{k+n} \}_{(1)} (L) 
= \{ l_{X},{\cal H}_{k} \}_{(2)}^{a,b} (L)
\end{equation}

\paragraph{Poisson maps}

Before turning to the study of the reductions of the KP hierarchies, 
let us exhibit some relations between the two quadratic brackets. We 
will use the invertible map in $\cal C$
\begin{equation}
p(L)=\partial^{-1}\tau(L)=D\partial^{-1}L^t\partial^{-1}\bar D.
\label{poimap}\end{equation}
Then a straightforward calculation leads to
\begin{equation}
\{ l_{X}\circ p,l_{Y}\circ p\}_{(2)}^a=-\{ l_{X},l_{Y}\}_{(2)}^b\circ 
p,
\end{equation}
which shows that \reff{pb2p} and \reff{pb2m} are equivalent Poisson 
brackets. However there is no relation between the hamiltonians
$\tr(L^{k\over n})$ and $\tr(p(L)^{k\over n-1})$.

There is another relation between the two brackets, which involves the
chiral superfield $\varphi$ satisfying the evolution equation 
\reff{fieq}.  Let us introduce the linear functional
$l_t=\int d^3{\underline x}(t\varphi)$, where $t({\underline x})$ is a Grassmann 
even superfield. We consider an enlarged phase space including  $\varphi$,
and extend the Poisson bracket \reff{pb2p} to this phase space by
\begin{equation}
\{ l_t,l_{Y} \}_{(2)}^a (L,\varphi)=\int d^3{\underline x} t((LY)_+.\varphi
+\Phi_Y\varphi),\,\,\,\{ l_{t},l_{t'} \}_{(2)}^a=0.
\end{equation}
Then one finds
\begin{equation}
\{ l_{X}\circ s,l_{Y}\circ s\}_{(2)}^a=\{ l_{X},l_{Y}\}_{(2)}^b\circ s,
\end{equation}
where the transformation $s$ has been defined in \reff{simil}. Notice 
that the hamiltonians are invariant functions for the transformation 
$s$, $\tr(L^{k\over n})=\tr(s(L)^{k\over n})$.

A last relation uses the antichiral superfield $\bar\chi$ satisfying 
the evolution \reff{chieq}.  Let us introduce the linear functional
$l_{\bar t}=\int d^3{\underline x}(\bar t\bar\chi)$, 
where $\bar t({\underline x})$ is a Grassmann 
odd superfield. We consider an enlarged phase space including  $\bar\chi$,
and extend the Poisson bracket \reff{pb2p} to this phase space by
\begin{eqnarray}
&\{ l_{\bar t},l_{Y} \}_{(2)}^a (L,\bar\chi)=\int d^3{\underline x} 
{\bar t}(-(LY)^t_+.\bar\chi
+\Phi_Y\bar\chi),&\\ &
\{ l_{\bar t_1},l_{\bar t_1} \}_{(2)}^a=-2\int d^{3}{\underline 
x}\bar t_1\bar\chi\bar\Phi[\bar t_2\bar\chi],&
\end{eqnarray}
where $\Phi$, $\bar\Phi$ are defined in equations (\ref{chipro},\ref{distri}).
Notice that this is a non-local Poisson bracket. One finds
\begin{equation}
\{ l_{X}\circ\sigma,l_{Y}\circ\sigma\}_{(2)}^a=-\{ 
l_{X},l_{Y}\}_{(2)}^b\circ\sigma,
\end{equation}
where the transformation $\sigma$ has been defined in \reff{chitra}.

\setcounter{equation}{0}
\section{Reductions of the KP hierarchy \label{reduc}}

In order to obtain consistent reductions of the KP hierarchy, we need to find Poisson 
submanifolds of the KP phase space. Considering first the  
quadratic  bracket \reff{pb2p}, we rewrite it as
\begin{eqnarray}
&\{ l_{X},l_{Y} \}_{(2)}^a (L) =\tr X\xi_{l_Y}^a,&\nonumber\\
&\xi_{l_Y}^a=(LY)_{+}L-L(YL)_{+}+\Phi_Y L+L\bar\Phi_Y.&
\label{hvf}\end{eqnarray}
$\xi_{l_Y}^a$ is the hamiltonian vector field associated with the function $l_Y$. One easily checks that if L has the form \reff{KPop}, then for any $Y$, 
$\xi_{l_Y}^a$ has the form $D(\sum_{i<n-1}\xi_i\partial^i)\bar D$.
It is obvious from \reff{hvf} that for any $Y$, if $L$ is in ${\cal C}_+$,
then $\xi_{l_Y}^a$ is also in ${\cal C}_+$. This means that the constraint
\begin{equation}L=L_+
\label{kdv}\end{equation}
 defines a Poisson submanifold. The hierarchies obtained in this way are 
the \n2 supersymmetric KdV hierarchies studied by Inami and Kanno
\cite{inami}, and the Lax operators \reff{kdv} already appeared in 
\cite{popo3}. The lowest order cases will be presented in the next section.
 
Another possible reduction is to take $L$ of the form
\begin{equation} L=L_++D\,\varphi\partial^{-1}\bar\varphi\bar D,\,\,\,\,\,
D\varphi=\bar D\bar\varphi=0.
\label{nls}\end{equation}
where $\varphi$ and $\bar\varphi$ are Grassmann even or odd chiral superfields.
With $L$ of the form \reff{nls} and $Y$ arbitrary, one finds
\begin{equation}
(\xi_{l_Y}^a)_-=D((LY)_+.\varphi+\Phi_Y\varphi)\partial^{-1}\bar\varphi
+\varphi\partial^{-1}(-(YL)_+^t.\bar\varphi+\bar\Phi_Y\bar\varphi))\bar D,
\end{equation}
Noticing that  $(LY)_+.\varphi$ is a chiral superfield and
$(YL)_+^t.\bar\varphi$ an antichiral superfield, it is easily checked that 
$\xi_{l_Y}^a$ is indeed tangent to the submanifold defined by the constraints \reff{nls}. 
It is possible to consider an enlarged phase space which coordinates are the fields in $L$
and $\varphi$, $\bar\varphi$. Let us introduce the linear functionals
\begin{equation}l_t=\int d^3{\underline x}(\varphi t),\,\,
l_{\bar t}=\int d^3{\underline x}(\bar t\bar\varphi),
\end{equation}
where $ t$ and $\bar t$ are general superfields, of the same 
Grassmann parity as $\varphi$ and $\bar\varphi$.
In this enlarged phase space, the second Poisson bracket, in the case when
$\varphi$ and $\bar\varphi$ are Grassmann even, is defined by \reff{pb2p}
and
\begin{eqnarray}
&\{ l_t,l_{Y} \}_{(2)}^a (L,\varphi,\bar\varphi)=\int d^3{\underline x} ((LY)_+.\varphi
+\Phi_Y\varphi)t, &\label{lfi}\\
&\{l_{\bar t},l_{Y} \}_{(2)}^a (L,\varphi,\bar\varphi)=\int d^3
{\underline x}\,\bar t(-(YL)_+^t.\bar\varphi
+\bar\Phi_Y\bar\varphi),&
\nonumber\end{eqnarray} and 
\begin{eqnarray}
&\{ l_t,l_{\bar t} \}_{(2)}^a (L,\varphi,\bar\varphi)=
\int d^3{\underline x}\, ({ L}_+.\bar t)t,&\label{bose}\\&
\{ l_{t_1},l_{t_2} \}_{(2)}^a=0,\,\,\,\,
\{ l_{\bar t_1},l_{\bar t_2} \}_{(2)}^a=0.&
\nonumber\end{eqnarray}
In the case when
$\varphi$ and $\bar\varphi$ are Grassmann odd, 
the last two lines should be modified to
\begin{eqnarray}
&\{ l_t,l_{\bar t} \}_{(2)}^a (L,\varphi,\bar\varphi)=
\int d^3{\underline x} (({ L}_+.\bar t)t-2\varphi t\Phi[\bar t\bar\varphi]),
\label{fermi}&\\&
\{ l_{t_1},l_{t_2} \}_{(2)}^a=2\int d^3{\underline x}\,\varphi t_1
\Phi[\varphi t_2],\,\,\,\,
\{ l_{\bar t_1},l_{\bar t_2} \}_{(2)}^a=-2\int d^3{\underline x}
\,\bar t_1\bar\varphi
\bar\Phi[\bar t_2\bar\varphi],&
\nonumber\end{eqnarray}
where the applications $\Phi$ and $\bar\Phi$ have been defined in
\reff{chipro}.
The lowest order case is $L=D(1+\varphi\partial^{-1}\bar\varphi)\bar D$. Then 
if $\varphi$ and $\bar\varphi$ are odd, the equation 
${d\over dt}L=[L^2_+,L]$ is the \n2 supersymmetric extension of the NLS
equation \cite{roelo}. The next-to-lowest order case is $L=D(\partial +H+\varphi\partial^{-1}\bar\varphi)\bar D$. 
If $\varphi$ and $\bar\varphi$ are even, the hamiltonian structure \reff{pb2p} reduces in 
this case to the classical version of the ``small'' $N=4$ superconformal algebra. Although 
the Poisson algebra contains $4$ supersymmetry
generators, the evolution equations \reff{KPeq} have only \n2 supersymmetry. 
This case was first obtained by another method which will be  
given, as part of a detailed study, in \cite{dgi}.

We now turn to the second quadratic bracket \reff{pb2m}. We rewrite it as
\begin{eqnarray}
&\{ l_{X},l_{Y} \}_{(2)}^b (L) =\tr X\xi_{l_Y}^b,&\nonumber\\
&\xi_{l_Y}^b=(LY)_{+}L-L(YL)_{+}+L{\lambda}(\Phi_Y)+{\bar\lambda}(\bar\Phi_Y)L.&
\label{hvfm}\end{eqnarray}
It is easily seen that neither the condition \reff{kdv}, nor the more complicated condition
\reff{nls} are admissible reductions in this case. The easiest way to 
find Poisson subspaces for the bracket \reff{pb2m} is to apply the 
map \reff{poimap} to the Poisson subspaces of the first quadratic 
bracket. From \reff{kdv}, we are then lead to the restriction:
\begin{equation}
L=L_++D\bar D\partial^{-1}H\partial^{-1}D\bar D
\label{a4}\end{equation}
With $L$ of the form \reff{a4} and $Y$ arbitrary, one finds
\begin{equation}
(\xi_{l_Y}^b)_-=D\bar D\partial^{-1}((LY)_+.H-(YL)_+^t.H+\res[L,Y]H)
\partial^{-1}D\bar D,
\end{equation}
which directly shows that condition \reff{a4} defines a Poisson submanifold for the 
Poisson bracket 
\reff{pb2m}. It turns out that \reff{a4} also defines a Poisson submanifold for 
the linear Poisson 
bracket \reff{pb1}. To show this we rewrite the linear bracket as
\begin{equation}\{ l_{X},l_{Y} \}_{(1)} (L) =\tr X\eta_{l_Y},\,\,\,
\eta_{l_Y}=[L,Y]_+-[L,Y_+]+{\lambda}(\Phi_Y)+{\bar\lambda}(\bar\Phi_Y).
\end{equation}
With $L$ of the form \reff{a4} and $Y$ arbitrary, one finds
\begin{equation}
(\eta_{l_Y})_-=D\bar D\partial^{-1}((Y_+-Y_+^t).H+\res[L,Y])\partial^{-1}D\bar D.
\end{equation}
Thus the reduced hierarchies defined by condition \reff{a4} are 
bi-hamiltonian. The lowest order cases will be studied in the next section.

Notice that the transformation \reff{simil} maps the systems satisfying the condition \reff{nls} with Grassmann even fields $\varphi$ and $\bar\varphi$
into systems satisfying condition \reff{a4} with 
\begin{equation}
H= \varphi\bar\varphi+\varphi^{-1}L_+.\varphi.
\end{equation}
Analogously, the transformation \reff{chitra} maps the systems satisfying the condition \reff{nls} with Grassmann odd fields $\varphi$ and $\bar\varphi$
into systems satisfying condition \reff{a4} with 
\begin{equation}
H= (-1)^n\left(\bar\varphi\varphi+(D\bar\varphi)^{-1}D(L_+^t.\bar\varphi)\right).
\end{equation}
Such transformations may be found in \cite{kriso,bokris}.

Finally we may consider the image of the Poisson subspace defined by 
\reff{nls} under the map $p$. One finds the condition
\begin{equation}
L=L_++D\bar D\partial^{-1}(H+\bar\varphi\partial^{-1}\varphi)\partial^{-1}D\bar D.
\label{n4}\end{equation}
The lowest order case is when $L_{+}=D\bar D$. The hamiltonian structure \reff{pb2m} 
reduces in 
this case to the classical version of the ``small'' $N=4$ superconformal algebra.
The equation ${d\over dt}L=[(L^{3})_{+}, L]$ becomes, after suitable 
redefinitions, the $N=4$ supersymmetric extension of the KdV equation 
derived in \cite{delivan} and written in \n2 superspace in 
\cite{dik}.

One can again consider an enlarged phase space which coordinates are 
the fields in $L$ and $\varphi$, $\bar\varphi$. The second quadratic bracket 
in this phase space is easily obtained from the first one by applying the map
$p$ to the first quadratic bracket. $p$ acts as the identity on $\varphi$ and 
$\bar\varphi$. As a consequence the Poisson brackets \reff{bose} and \reff{fermi} keep the same form, whereas \reff{lfi} should be modified to
\begin{eqnarray}
&\{ l_t,l_{Y} \}_{(2)}^b (L,\varphi,\bar\varphi)=\int d^3{\underline x} (\res\left(\tau((YL)_+)\lambda(\varphi)\right)
+\Phi_Y\varphi)t, &\\
&\{l_{\bar t},l_{Y} \}_{(2)}^b (L,\varphi,\bar\varphi)=\int d^3
{\underline x}\,\bar 
t(-\res\left(\bar\lambda(\bar\varphi)\partial^{-1}\tau((LY)_+)\partial\right)
+\bar\Phi_Y\bar\varphi).&
\nonumber\end{eqnarray}

\setcounter{equation}{0}
\section{Examples and comparison with other works \label{examples}}

This paragraph is devoted to the presentation of the simplest integrable equations obtained using our formalism.

Considering first the condition \reff{kdv}, the simplest example is the lax operator $L = D (\partial + W) \bar{D}$. Then the
evolution equation
\begin{equation}
{d\over dt}L=[L^{3\over 2}_+,L],
\end{equation}
leads to the equation
\begin{equation}
8\partial_{t} W = 2W_{xxx}+6\left( (DW)(\overline{D}W) \right)_{x}- 
\left( W^3 \right)_{x},
\end{equation}
which coincide after the redefinition $W=2i\Phi$ with the $a=-2$ \n2 
extension of the KdV equation in the classification of 
Mathieu \cite{Mathieu1,math2}. The Lax operator given in \cite{Mathieu1} may be
obtained from $L$ in the following way. Let us consider the operator
\begin{equation}
L_{-2} = L+L^t=\partial^2+W[D,\bar D]+(DW)\bar D-(\bar D W)D.
\label{eqam2}\end{equation}
$L$ is in $\cal C$ and $L^t$ is in $\bar{\cal C}$. If we remember that the product of an element in $\cal C$ and an element in $\bar{\cal C}$ always vanishes, we immediately get that a square root of $L_2$ with highest derivative term equal to $\partial$ is $(L_{-2})^{1\over 2}=L^{1\over 2}-(L^{1\over 2})^t$. From this we deduce the relation $(L_{-2})^{3\over 2}=L^{3\over 2}
-(L^{3\over 2})^t$. As a consequence $L_{-2}$ satisfies the evolution equation
\begin{equation}{d\over dt}L_{-2}=[L^{3\over 2}_+,L]+([L^{3\over 2}_+,L])^t
=[(L_{-2})^{3\over 2},L_{-2}],\end{equation}
which is thus an equivalent Lax representation for equation \reff{eqam2}. 
 
As the next example, we consider the Lax operator 
\begin{equation}
L=D(\partial^2+V\partial+W)\bar D
\end{equation}
Then the evolution equation ${d\over dt}L=[L^{2/3}_{+},L]$ should coincide, after suitable redefinitions, with one of the three \n2 supersymmetric extensions of the Boussinesq equations derived in \cite{Ivanov1}. Indeed
one can check that the Lax operator they give for the $\alpha = -1/2$
equation may be written as
$L^{(1)}=L+\bar D\partial^2 D$. Then one easily obtains 
$(L^{(1)})^{2\over 3}=L^{2\over 3}+\bar D\partial D$, and the evolution equation for $L^{(1)}$ is easily deduced from that of $L$
\begin{equation}
{d\over dt}L^{(1)}={d\over dt}L=[(L^{(1)})^{2\over 3},L^{(1)}].
\end{equation}

Turning now to condition \reff{a4}, the lowest order case corresponds to
the Lax operator 
$L=D\bar D+D\bar D\partial^{-1}W\partial^{-1}D\bar D$. Then the 
equation ${d\over dt}L=[(L^{3})_{+}, L]$ becomes, after suitable 
redefinitions, the \n2 supersymmetric extension of the KdV equation 
with parameter $a=4$,
\begin{equation}
\partial_{t}W = W_{xxx} + \frac{3}{2}\left( [ D ,\overline{D} ] W^2 \right)_{x} - 3\left( (DW)(\overline{D}W) \right)_{x} + (W^3)_{x}
\end{equation}
Notice that, all integer powers of $L$ define 
conserved charges in this case (an alternative Lax operator with the 
same property was derived in \cite{krisoto}).

The last example that we shall study is the Lax operator  
\begin{equation}
L = D \left( \partial + V \right) \overline{D} + D\overline{D}\partial^{-1}W\partial^{-1}D\overline{D}.
\end{equation}
Then the equation
\begin{equation}
\partial_{2} L = [L_{+},L] 
\end{equation}
explicitely reads
\begin{eqnarray}
\partial_{2} V &=& 2 W_{x}\\
\partial_{2} W &=& [ D, \overline{D} ]W_{x} +VW_{x} +(DV)( \overline{D}W)+( \overline{D}V)(DW).
\end{eqnarray}
This equation is identical, up to a rescaling of time, to the \n2 supersymmetric extension 
of the Boussinesq equation with parameter $\alpha = -2$
derived in \cite{Ivanov1}.

\setcounter{equation}{0}
\section{From $N=2$ to $N=1$ superspace \label{n1susy}}

$N=2$ extensions of the KP and KdV hierarchies
have been studied in several articles
\cite{inami,ghosh,dasb1,dasp} using an $N=1$ superspace formalism. In this section we wish to relate the KP 
hierarchies that we described in section \ref{main} to those given in
the litterature. The first step will be to relate our $N=2$ algebra ${\cal C}$
of pseudo-differential operators to the $N=1$ algebra of pseudo-differential operators. 
An operator $L=D{\cal L}\bar D$ in ${\cal C}$ should be considered as acting on a chiral 
object $\Psi$, $D\Psi$=0, and this action writes
\begin{equation}
L .\Psi=D{\cal L}\bar D .\Psi={\cal L}\partial .\Psi+(D .{\cal L})\bar D .\Psi.\label{R1}
\end{equation}
We shall use the following combinations of the chiral derivatives
\begin{equation}
D_1=D+\bar D,\,\, D_2=-D+\bar D,\,\, D_1^2=-D_2^2=\partial,\,\,
\{ D_1,D_2\}=0.\end{equation}
Then the action of $L$ on $\Psi$ is
\begin{equation}
L .\Psi=({\cal L}\partial+(D .{\cal L})D_1) .\Psi.
\end{equation}
We then choose to associate to the $N=2$ pseudo-differential operator $L$
the $N=1$ pseudo-differential operator $\underline L$ given by
\begin{equation}
{\underline L}={\cal L}\vert_{\theta_2=0}\partial+
(D .{\cal L})\vert_{\theta_2=0}D_1.
\end{equation}
It is easily checked that this correspondence respects the product,
$\underline{LL'}=\underline{L}\,\,\underline{L'}$. It also has the property
\begin{equation}
\underline{L_+}={\underline L}_{>0}.
\end{equation}
That is to say that the image of an $N=2$ differential operator is a strictly differential 
$N=1$ operator, without the non-derivative term. Notice also the useful relations
\begin{eqnarray}
& \res (\underline{L})=(D. {\res}(L))\vert_{\theta_2=0},\,\, &\\ 
&\tr(L)=\tr(\underline{L})\equiv
\int d^2{\underline x}  {\res}(\underline{L}),\,\,\,\int 
d^2{\underline x}\equiv\int dxd\theta_1&
\end{eqnarray}
where the residue of the operator $\underline L$ is the coefficient of 
$D_1^{-1}\equiv D_1\partial^{-1}$. From now on, all expressions will be written in $N=1$ 
superspace, and we drop the index of $D_1$ and $\theta_1$. The KP hierarchy described in 
section \ref{main} may be described in $N=1$ superspace as follows. We consider an operator
$\underline L$ of the form
\begin{equation}
\underline{L}=D^{2n}+\sum_{p=1}^\infty w_pD^{2n-p-1}
\end{equation}
and consider evolution equations
\begin{equation}
{\partial\over\partial t_k}\underline{L}=[\underline{L}^{k\over n}_{>0},L]
\label{R9}\end{equation}
This is nothing but the non-standard supersymmetric KP hierarchy described in \cite{ghosh,dasb1}. The evolution equations (\ref{R9}) admit
the conserved quantities $H_p=\tr(\underline{L}^{p\over n})$, and they are bi-hamiltonian. 
The first Poisson bracket is easily deduced from its $N=2$
counterpart (\ref{pb1}). With 
$l_{\underline{X}}=\tr(\underline{L}\,\,\underline{X})$, we have
\begin{equation}
\{ l_{\underline{X}},l_{\underline{Y}}
 \}_1=\tr L([\underline{X}_{>0},\underline{Y}_{>0}]-
[\underline{X}_{\leq 0},\underline{Y}_{\leq 0}])
\end{equation}
As in the $N=2$ formalism, this is a standard bracket associated with a non-antisymmetric 
$r$ matrix. As a consequence, the two quadratic brackets 
deduced from (\ref{pb2p}) and \reff{pb2m} are
quite complicated. They involve the quantity $\psi_{\underline{X}}$ defined up to a constant by
$D\psi_{\underline{X}}= {\res}[{\underline{L}}\, ,{\underline{X}}]$. The first one is
\begin{eqnarray}
&\{ l_{\underline{X}},l_{\underline{Y}}
 \}_2^a(L)=\tr(\underline{L}\,\,\underline{X}(\underline{L}\,\,\underline{Y})_+
-\underline{X}\,\,\underline{L}(\underline{Y}\,\,\underline{L})_+)
+\int d^2{\underline x} (-\psi_{\underline{Y}}\, {\res}[{\underline{L}},{\underline{X}}]\nonumber&\\& 
+{\res}[{\underline{L}}\, ,{\underline{Y}}]\,
 {\res}(\underline{X}\,\,\underline{L}\,D^{-1}) 
- {\res}[{\underline{L}}\, ,{\underline{X}}]\,
 {\res}(\underline{Y}\,\,\underline{L}\,D^{-1})
).&
\end{eqnarray}

The Poisson bracket \reff{pb2m} becomes
\begin{eqnarray}
&\{ l_{\underline{X}},l_{\underline{Y}}
 \}_2^b(L)=\tr(\underline{L}\,\,\underline{X}(\underline{L}\,\,\underline{Y})_+
-\underline{X}\,\,\underline{L}(\underline{Y}\,\,\underline{L})_+)
+\int d^2{\underline x} (\psi_{\underline{Y}}\, {\res}[{\underline{L}},{\underline{X}}] \nonumber&\\& 
+{\res}[{\underline{L}}\, ,{\underline{Y}}]\,
 {\res}(\underline{L}\,\,\underline{X}\,D^{-1}) 
- {\res}[{\underline{L}}\, ,{\underline{X}}]\,
 {\res}(\underline{L}\,\,\underline{Y}\,D^{-1})
),&
\end{eqnarray}
and already appeared in  \cite{dasp}. 
It is not a difficult task to obtain the $N=1$ restrictions which correspond to the \n2 conditions (\ref{kdv},\ref{nls},\ref{a4},\ref{n4}). Some of the lax 
operators obtained in this way are already known, in particular those satisfying \reff{kdv} from \cite{inami} and  the lowest order operator coming from
\reff{nls} with odd $\varphi$ and $\bar\varphi$, which is the super-NLS Lax operator obtained in \cite{dasb1}.

\setcounter{equation}{0}
\section{Conclusion}
 
An easy generalization of the hierarchies presented in this article would be to consider multi-components KP hierarchies, that is to say replace the fields 
$\varphi$ and $\bar\varphi$ in \reff{nls} and \reff{n4} by a set of $n+m$ fields
$\varphi_i$ and $\bar\varphi_i$, $n$ of them being Grassmann even and the other 
$m$ being Grassmann odd. 
For the lowest order case of equation \reff{nls}, such a generalization has been considered in \cite{bokris}. The Lax representation that we propose for such hierarchies has the advantage that one does not need to modify the definition of the residue. For the next to lowest order case of equation \reff{nls}, and
the lowest order case of equation \reff{n4}, it should be possible to obtain
in this way hierarchies based on $\cal W$-superalgebras with an arbitrary number of supersymmetry charges.

Little is known about the matrix Lax formulation of the hierarchies presented here. In the case of operators satisfying 
condition \reff{kdv},
such a matrix Lax formulation was constructed in $N=1$ superspace
by Inami and Kanno \cite{inami,ina1}. It involves the loop superalgebra based on $sl(n\vert n)$. What we know about the matrix Lax formulation in \n2 superspace for hierarchies based on Lax operators satisfying conditions \reff{kdv} or \reff{nls} will be reported elsewhere. Notice that we obtained the form \reff{kdv} of the scalar Lax operators from a matrix Lax representation, and only later became aware of reference \cite{popo3} where these operators also appear.


\end{document}